\documentclass[twocolumn,showpacs]{revtex4}
\usepackage{graphicx}
\usepackage{amsmath}
\usepackage{latexsym}

\begin{document}

\title{Deviations from Full Aging in Numerical Spin Glass Models} 
\author{Jesper Dall}
\author{Paolo Sibani}\email[Corresponding author ]{paolo@planck.fys.sdu.dk}
\affiliation{Fysisk Institut, Syddansk Universitet, Campusvej 55, 
DK--5230 Odense M, Denmark} 

\date{\today}

\begin{abstract}
{\small 
The deviations from full or pure aging behavior, i.e.\ perfect 
$t/t_w$ scaling of the correlation and response functions of 
aging glassy systems, are not well understood theoretically. 
Recent experiments of Rodriguez 
et al.~(Phys.\ Rev.\ Lett.\ \textbf{91}, 037203 (2003)) have 
shown that full aging applies to the thermoremanent 
magnetization in the limit of infinite cooling rate during 
the initial quench. In numerical models, instantaneous 
thermal quenches can be---and usually are---applied. 
This has motivated the present numerical investigations of the
aging behavior of Edwards-Anderson Ising spin glass models 
with both short- and long-range Gaussian interactions, 
respectively.
We sample the distribution of residence time $t$ in suitably 
defined metastable valleys entered at age $t_w$, finding that 
the deviations from $t/t_w$ scaling are small and decrease 
systematically as the system size grows and/or the temperature 
decreases. Finally, the connection between this behavior and 
the scaling of the correlation function itself is discussed.
}
\end{abstract}

\maketitle

\section{Introduction} \label{introduction}
In a typical aging experiment a spin glass
sample is quenched below the glass transition temperature 
in a small magnetic field.
At time $t_w$ after the quench, the field is switched off and the
subsequent decay of the thermoremanent magnetization is measured.
It is well known that linear response functions
in spin glasses strongly depend on the age or
waiting time $t_w$. In general, the rate of change of macroscopic
averages in glassy systems systematically decreases with $t_w$.

Considering the broad variety of aging systems, microscopic time 
scales are hardly significant, which highlights $t/t_w$ as the
obvious choice of scaling variable for aging data. The
data collapses obtained in this fashion are fairly 
good but never perfect: In spin-glass linear response 
experiments~\cite{vincent97,bouchaud99} much better scalings 
are obtained using $t/t_w^\mu$, with the exponent
$\mu$ slightly smaller than one.\ This situation
is referred to as sub-aging, since the apparent age $t_w^\mu$
is shorter than $t_w$, as opposed to `pure' or `full'
$t/t_w$ aging. In gels~\cite{bissig03} and
colloidal glasses~\cite{viasnoff02},
the distribution of the 
waiting time between pairs of consecutive `anomalous
fluctuations' features the opposite deviation, i.e.
super- or hyper-aging, where $\mu > 1$. 
Both sub- and super-aging pose a theoretical challenge as
they require the introduction and interpretation
of an additional time scale. This remains
true even when $\mu$ is very close to unity, as
in the present study.

Qualitatively, aging reflects the trapping of 
trajectories in metastable `valleys' of ever increasing
thermal stability~\cite{Sibani89,Bouchaud92,sibanidall03a}:
the pseudo-equilibrium fluctuations
occur for $t < t_w$ within the valley selected at 
time $t_w$, while the exploration of new valleys
occurs for $t> t_w$ and produces
the non-equilibrium part of the aging dynamics.
In this picture, the spin rearrangements
leading from one valley to the next, `quakes'
in our terminology, would appear as
anomalous events in a fluctuating background. 
The waiting time between quakes corresponds to
the valley residence time, whose
statistical properties have repeatedly been
considered in models of complex 
relaxation~\cite{Sibani87,Bouchaud92,rinn00}. 
According to a recent description of non-equilibrium 
dynamics as a log-Poisson process~\cite{sibanidall03a}, 
the distribution $R(t \mid t_w)$ of
residence time $t$ in valleys entered at
time $t_w$ should scale as $t/t_w$
and possess a finite average residence
time proportional to $t_w$. The best obtainable data 
collapse of the corresponding simulational 
data reveals however a slight super-aging behavior.

On the experimental side, it has recently 
been shown~\cite{Rodriguez03} that sub-aging of 
the thermoremanent magnetization can be strongly suppressed
by increasing the cooling rate of the quench.
For the instantaneous quench implied
in isothermal numerical simulations with a
random initial configuration, this 
strengthens the expectation that full aging
applies, a conclusion which is not inequivocally supported
by the existing numerical studies
of the correlation function~\cite{kisker96,berthier02}.

The paper is organized as follows: In Section~\ref{MethodModels}
we summarize the exploration method used to
identify the metastable valleys and describe the procedures
followed. In Section~\ref{NumericalResults} we
present our numerical findings for the scaling of 
the residence time distribution in these valleys 
for spin glass models 
defined on Euclidean lattices with nearest neighbor 
interactions as well as on random regular graphs. 
In particular, we focus on the system size and 
temperature dependence of $\mu$.
Section~\ref{Discussion} compares 
the empirical properties of this distribution with the
prediction of the idealized log-Poisson theory
of Ref.~\cite{sibanidall03a}. 
We show that the theoretical description becomes
gradually better
in the sense that full aging is slowly approached 
as the temperature decreases or the system size increases.
Finally, the implications for the correlation function
are discussed, and tentative conclusions 
are drawn regarding the geometrical
origin of the deviations from pure aging behavior.

\section{Method and Models} \label{MethodModels}
While the fluctuation dynamics of glassy systems within
metastable valleys is uncontroversially
determined by the local free energy, 
a similarly well established theoretical framework 
for the irreversible quakes interrupting
the pseudo-equilibrium regime is lacking.
Surprisingly, there has not been much focus 
on how these jumps or quakes should be identified and
characterized. This question is subtle since the 
first occurrence of a certain rearrangement
could well have a non-equilibrium
character, while subsequent rearrangements
of similar nature might represent
pseudo-equilibrium fluctuations within a larger valley.

The classic analysis of energy 
landscapes~\cite{Stillinger84,doliwa03b} 
samples local energy minima using e.g.\ downhill
search algorithms as thermal quenches, and is
not overly concerned with how attractors are dynamically
selected in the course of unperturbed isothermal aging.
An empirical procedure to identify the 
quakes~\cite{sibanidall03b} distinguishes `first' and 
`subsequent' occurrences of certain events by keeping 
track of the sequence of energies visited while aging.
Within this sequence, states of energy lower than all 
previously visited states, together with 
barriers~\footnote{Barriers are defined as energy 
differences with respect to the current lowest energy state.} 
higher than all previously surmounted barriers, define a 
quake as a rearrangement which overcomes a barrier 
of record height \emph{and} leads to a state of energy 
lower than the lowest energy seen so far. A valley is then 
simply the set of states visited between two subsequent 
quakes, where onset of the latter is marked by the 
barrier records. 
Very restrictive by construction, this procedure yields 
no output if the dynamics is time translationally 
invariant, and is therefore not available to analyze 
e.g.\ the diffusive regime of supercooled glass
formers. Within the aging regime, it produces a highly 
non-trivial yet relatively simple picture of the energy 
landscape of spin glasses~\cite{dall03a}.

The Edwards-Anderson~\cite{edwards75} model
considered below is an archetypal glassy system
which possesses a complex dynamics while
remaining relatively easy to simulate.
The energy of a configuration in this model is given by
\begin{equation}
E(\{s_1,\ldots,s_N\}) = - \frac{1}{2}\sum_{i,j} J_{ij} s_i s_j.
\label{SGenergy}
\end{equation}
The couplings $J_{ij}$ are symmetric, independent Gaussian
variables of unit variance. We used standard cubic
lattices with periodic boundary conditions, where
$J_{ij}$ is non-zero if and only if $i$ and $j$ are nearest neighbors.
To investigate more general networks as well, we
additionally explored $k$-regular random graphs,
i.e.\ where each spin interacts with exactly $k$ other spins chosen
at random. This is similar to the Viana-Bray model~\cite{viana85}, except
that the number of links emanating from each spin was fixed to $k=6$
for better comparison with the Euclidean $3d$ case.

The systems considered contain up to
$N=40^3$ spins in $3d$,
and $N=16000$ spins in the regular random graph case.
The dynamics is simulated using the Waiting Time
Method (WTM)~\cite{Dall01}.
This rejectionless method is equivalent
to the Metropolis algorithm, but much faster at low $T$
for simulations of glassy systems.
The intrinsic, size independent time variable $t$ of the WTM
corresponds to the number of Monte Carlo (lattice)
sweeps in the Metropolis algorithm as well as to the physical time of a real
experiment.

\section{Numerical Results} \label{NumericalResults}
\begin{figure}[t]
\begin{center}
\includegraphics[width=8.5cm]{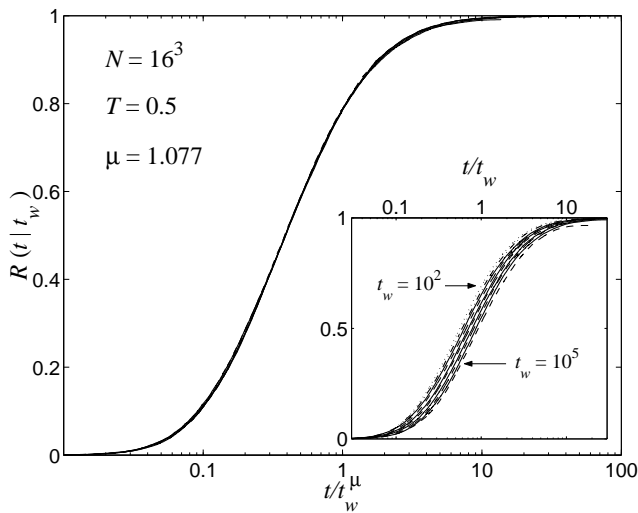}
\caption{ \small
The probability distribution of residing at most a time $t$
in a valley entered at time $t_w$ is calculated for
a set of $10$ waiting times
equally spaced on a logarithmic scale in the interval
$10^2 \leq t_w \leq 10^5$. Each distribution function is 
plotted versus the scaled variable $t/t_w^\mu$, with 
$\mu = 1.077$, producing the perfect data collapse shown 
in the main panel. All data pertain to a $3d$ spin glass 
as defined in Section~\ref{MethodModels}.
The insert shows that choosing $\mu =1$ gives a visible 
spread of the curves, the latter being shifted to the right 
as $t_w$ grows. The apparent age $t_w^\mu$ inferred from
the scaling plots is thus larger than the actual age,
which is super-aging behavior.
}
\label{L16}
\end{center}
\vspace{-0.5cm}
\end{figure}

Using the valley-identification method outlined in 
Section~\ref{MethodModels}, we examine the distribution 
of residence time $t_{res}$ in a valley entered at $t_w$, i.e.\ 
\begin{equation} 
R(t \mid t_w) = {\rm Prob}( t_{res} < t \mid t_w ).
\label{R}
\end{equation}
The distribution for a $3d$ system with $N=16^3$ spins
at temperature $T=0.5$ is shown in Fig.~\ref{L16} 
for a wide range of $t_w$ values.
To collapse the data shown we used the scaling variable
$t/t_w^{\mu}$ with $\mu=1.077(6)$. The insert confirms 
that full aging, $ \mu =1$, is only a fair approximation.
The data collapse thus obtained
is closely matched by the graph of the function
$f(x) = 1 - (1+x)^{-\alpha}$, 
where $\alpha \simeq 2$.
The latter is not shown in order to avoid masking 
the data---see Fig.~3 of 
Ref.~\cite{sibanidall03a} for a direct comparison.

\begin{figure}[t]
\begin{center}
\includegraphics[width=8.5cm]{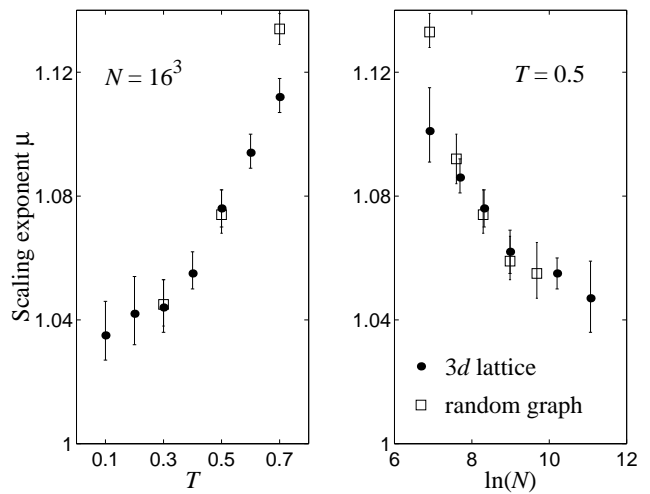}
\caption{ \small 
The scaling parameter $\mu$ as a function of temperature (left panel) 
and system size (right panel). For each data point, the value of $\mu$ 
is estimated as in Fig.~\ref{L16}. The figure shows that super-aging 
($\mu > 1$) fades away for low temperatures and large system sizes, 
indicating that deviations from full aging stem primarily 
from finite size effects.
}
\label{mu}
\end{center}
\vspace{-0.5cm}
\end{figure}

The present simulations greatly extend our earlier 
findings~\cite{sibanidall03a,dall03a} by probing the 
size and temperature dependence of the small deviations 
from unity of the scaling exponent $\mu$.
Numerical investigation of such effects requires the
removal of any visible statistical flutter. This has 
been achieved by averaging over at least 4,000 runs with 
different realizations of the couplings $J_{ij}$ for every 
combination of $N$ and $T$. The calculations were performed on
a large Linux cluster, with a total run-time equivalent to
four years of computation on a single 2GHz processor.
The result of these simulations is summed up in Fig.~\ref{mu}.

The high quality of the data collapse in Fig.~\ref{L16} 
is representative for all sets of runs at fixed size 
and temperature on which the data points in Fig.~\ref{mu} 
are based. This suggests that the `best' value of $\mu$ 
can readily be estimated by eye, as small variations of 
$\mu$ immediately produce a visible scatter. To remove 
any personal bias, we did however utilize a quantitative
error measure $e(\mu)$ gauging the deviation from perfect
collapse of a set of scaled exit time distributions:
The horizontal standard deviation for a set of $R(t \mid t_w)$ 
values chosen equidistantly between 0.2 and 0.8 is 
calculated for ten values of $t/t_w^\mu$ at fixed $\mu$. 
The value of $e(\mu)$ is the sum of these 
standard deviations, and it attains a minimum within a 
`reasonable' $\mu$-interval. The uncertainty bracket around 
the $\mu$ value producing the least scatter is arbitrarily 
defined by the two values of $\mu$ for which $e(\mu)$ is no 
more than $20$\% larger then its minimum value. 

Figure~\ref{mu} provides information on 
$\mu=\mu(N,T)$ in both $3d$ lattices ($\bullet$) and 
random graphs ($\Box$). The data points and error bars
are determined as explained above, 
ensuring that the quality of the data collapse 
behind the estimate of $\mu$ is near-perfect, 
i.e.\ as good as in Fig.~\ref{L16}.
In the left panel, $\mu(N,T)$ is shown as a function 
of $T$ for fixed size $N=16^3$.
There is a clear decreasing trend, which however seems 
to taper off as the temperature decreases. The diminishing 
deviation from pure scaling concurs with the results by 
Kisker et al.~\cite{kisker96} regarding
the correlation function $C(t_w,t_w+t)$ in $3d$ Gaussian 
spin glasses. The right panel illustrates the behavior
of $\mu(N,T)$ for $T=0.5$ as a function of the logarithm
of $N$. Again, as the number of degrees of freedom grows, 
super-aging is suppressed, albeit in a very slow fashion. 
Finally, we note the striking similarity between 
the Euclidean and random networks data in Fig.~\ref{mu}.
The scaling of the residence time distribution seems to be
indifferent to the topology of the graph considered.

When the system size becomes too small, the quality of 
the scaling of $R(t \mid t_w)$ begins to deteriorate. 
This applies to systems defined on $3d$ lattices
as well as on random graphs. A possible explanation is a 
finite time (or size) effect: by construction, the valley 
containing the ground state will never be exited. By the 
same token, very low lying valleys, which are most likely 
entered for large $t_w$ and in small systems, are never 
exited during the simulation time. This flattens the 
s-shape of the distribution for large $t_w$ values and 
ruins the scaling. Lacking precise estimates of $\mu$ for 
$N < 1000$, we present data for larger systems only.

\section{Discussion} \label{Discussion}
The residence time analyzed in the previous section
appears from the outset conceptually similar to the trapping time 
introduced in a heuristic trap model by Bouchaud~\cite{Bouchaud92}
and later considered by Rinn et al.~\cite{rinn00} 
in a development specifically dealing 
with sub-aging behavior. However, the statistical properties 
hypothesized in the above models deviate considerably 
from those presently found. Crucially, the fitting form
\begin{equation}
R(t \mid t_w) = 1 -(1 + t/t_w^\mu)^{-\alpha}
\label{fit}
\end{equation} 
with $\alpha =2$ yields an average residence time 
\begin{equation}
\langle t_{res} \rangle 
= \int \frac{d R}{d t} t dt 
= \frac{t_w^\mu}{\alpha -1} 
= t_w^\mu,
\end{equation} 
which is finite, in contrast with the assumption
of the trap model~\cite{Bouchaud92}. Secondly, the
only temperature dependence is the
weak increase of $\mu$, while the exponent $\alpha$
remains independent of $T$.

The numerical results of Fig.~\ref{mu} points to pure 
aging as the correct limit for small $T$ or large $N$, 
although the sub-logarithmic size dependence of $N$ 
prevents us from turning this into a firm conclusion. 
The possible physical origin of the deviations from 
$\mu=1$ for finite $N$ was already hinted to: Eventually, 
the trajectory enters the valley containing the ground 
state and never leaves. While this happens at extreme 
time scales for realistic system sizes, fewer valleys
will in general be available as the trajectories approach 
the ground state. The residence time for valleys entered 
at late stages will hence tend to be longer than 
otherwise expected. In accord with this interpretation, 
lowering $T$ has an effect on $\mu$ similar to the effect 
of enlarging $N$. In any case, the deviation from a pure 
log-Poisson description of quake 
dynamics~\cite{sibanidall03a} becomes very small for 
large $N$, irrespective of the topology of the system.
This supports the generality of 
log-Poisson statistics, whose prediction for the 
distribution of the residence time in a valley 
entered at age $t_w$ coincides with Eq.~(\ref{fit}) 
for $\mu=1$~\cite{sibanidall03a}. 

While $R(t \mid t_w)$ might be directly accessible in 
some experimental situations, its relationship to 
correlation functions for spin glasses is indirect and 
requires further assumptions and/or empirical input. If 
one assumes with Ref.~\cite{rinn00} full decorrelation 
once a trap is left, and that the age $t_w$ uniquely 
determines the properties of the `initial' trap, the 
correlation function basically coincides with the 
probability that no events occur from $t_w$ to $t_w +t$.
As the latter is again given by Eq.~(\ref{fit}),
this would allow us to extend the conclusions
already drawn for $R$ to the correlation function.

To assess the validity of such reasoning we will use 
a cartoon picture of the quake dynamics which neglects 
the internal structure of the valleys. This is a 
questionable step, as the time spent searching for 
the bottom state within a valley is of the same order 
of magnitude as the residence time itself~\cite{dall03a}, 
i.e.\ quakes are not instantaneous 
events. With this caveat in mind, we write the 
non-equilibrium thermal correlation function as an average 
over the distribution of quakes~\cite{sibanidall03a}:
\begin{equation}
C(t_w, t_w +t) = 
\sum_{m,k} P_m(t_w) P_k(t_w, t_w+t) c(m,m+k), 
\label{C3}
\end{equation}
valid in the $t > t_w$ regime.
In the above formula, $P_m(t_w)$ is the probability 
that $m$ quakes are recorded during the aging time $t_w$, 
$P_k(t_w, t_w+t)$ is the probability that $k$ quakes 
occurred between $t_w$ and $t+t_w$, and, finally, 
$c(m,m+k)$ is the configurational overlap between the 
`bottom states' of the $m$'th and $(m+k)$'th valley.

The log-Poisson statistics prescribes the form of both $P_m(t_w)$
and $P_k(t_w, t_w+t)$~\cite{sibanidall03a}. While the 
latter scales with $t/t_w$, the former depends on 
$t_w$ alone. This ruins pure aging for $C(t_w, t_w +t)$ unless the 
sum of $P_m(t_w)$ over $m$ happens to
factor out. This can only occur if the overlap
$c(m,m+k)$ does not depend on $m$. Perfect invariance of $c(m,m+k)$
with respect to the index $m$ must fail at some stage, since, again, 
the quake-induced rearrangements
are likely to change as the ground state is approached. 
Numerical investigations presented elsewhere~\cite{sibanidall03a,dall03a} 
show that $c(m,m+k)$ has a small but systematic 
$m$ dependence already for small $m$ values. This excludes 
the possibility of `perfect' pure aging of the correlation function 
for the numerical models studied, even though pure aging is 
well fulfilled for the waiting time distribution itself.
 
As the experimental results of Rodriguez
et al.~\cite{Rodriguez03} clearly support pure aging of
the correlation function, it is possible
that a description more refined than what Eq.~(\ref{C3})
provides would modify our conclusion.
In view of the smallness of the effects involved, it is
also possible that some subtle dynamical feature
of the experimental systems is missed by the computer models
in the size range considered. A final possibility, which
we regard as the most likely, is that
$c(m,m+k)$ is indeed independent of $m$ for systems in the
macroscopic limit. 
This would emphasize the extreme non-equilibrium nature of 
the observed low $T$ dynamics in spin glasses.

\vspace{0.5cm}
\noindent {\bf Acknowledgments}:
The  Danish Center for Scientific Computing has
provided generous time on the Horseshoe Linux cluster
in Odense. The authors are also indebted to Greg Kenning 
and Stefan Boettcher for discussions. 
This project has been supported by Statens Naturvidenskabelige 
Forskningsr{\aa}d.
\bibliographystyle{unsrt}

\bibliography{SD-meld,sg}
\end{document}